\documentclass[preprint2]{aastex}

\shorttitle{The metallicity spread of $\omega$ Cen.}
\shortauthors{Villanova et al.}

\begin{document}

\title{The metallicity spread and the age-metallicity relation of $\omega$ Centauri \thanks{Based
    on FLAMES+GIRAFFE@VLT  observations under
    the  program  082.D-0424(A)}}


\author{S. Villanova and D. Geisler}
\affil{Departamento de Astronomia, Casilla 160, Universidad de Concepcion, Chile}
\email{svillanova@astro-udec.cl}

\author{R.G. Gratton}
\affil{INAF - Osservatorio Astronomico di Padova, Vicolo dell’Osservatorio 5, 
35122 Padova, Italy }

\and

\author{S. Cassisi}
\affil{INAF - Ossservatorio Astronomico di Teramo, via M. Maggini, 64100
Teramo, Italy}




\begin{abstract}
$\omega$ Centauri is a peculiar Globular Cluster formed by a complex stellar
population. To shed light on this, we studied 172 stars belonging to the 5 SGBs
that we can identify in our photometry, in order to measure their [Fe/H]
content as well as estimate their age dispersion and the age-metallicity
relation. 
The first important result is that all of these SGBs has a distribution 
in metallicity with a spread that exceeds the observational errors and 
typically displays several peaks that indicate the presence of several sub-populations.
We were able to identified at least 6 of them based on their mean [Fe/H]
content. These metallicity-based sub-populations are seen to varying extents
in each of the 5 SGBs.

Taking advantage of the age-sensitivity of the SGB we showed that, first of all,
at least half of the sub-populations have an age spread of at least 2 Gyrs.
Then we obtained an age-metallicity relation that is the most complete up to
date for this cluster.

The interpretation of the age-metallicity relation is not straightforward, but
it is possible that the cluster (or what we can call its progenitor) was
initially composed of two populations having different
metallicities. Because of their age, it is very unlikely that the
most metal-rich derives from the most metal-poor by some kind of chemical
evolution process, so they must be assumed as two independent primordial
objects or perhaps two separate parts of a single larger object, that
merged in the past to form the present-day cluster. 
\end{abstract}


\section{Introduction}

\begin{figure*}[ht!]
\epsscale{1.50}
\plotone{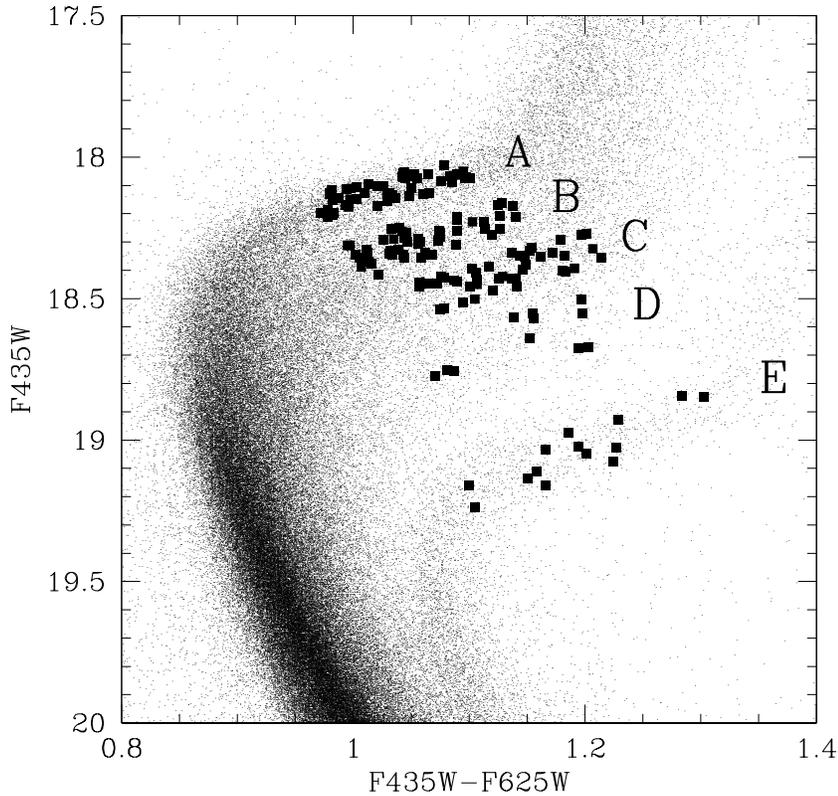}
\caption{SGB region of $\omega$ Cen with the 5 branches 
  identified in HST data \citep{vi07} and the target stars indicated.}
\label{f1}
\end{figure*}

Omega Centauri is a fascinating and enigmatic object:\ it appears to be a
globular cluster (GC), but it has a very complex stellar population, and
with its unusual mass ($M\sim3\times10^6M_{\odot}$) it has often been
suggested to be the remains of a larger stellar system.  It has
received a large amount of attention; for a review see \citet{me03}.
One of the most interesting results \citep{be04} 
was the discovery that over a range of at least two
magnitudes the main sequence splits into red and blue branches.
Follow-up spectroscopic studies at medium resolution led to 
the finding that, contrary to any expectation from canonical stellar models,
the bluer branch of the MS is more metal-rich than the red \citep{pi05}.  At the moment, the only
explanation of the photometric and spectroscopic properties of the
double main sequence that is at all plausible is that the bluer branch
of the MS has an unusually high helium content \citep{no04,ki12}.

\begin{figure}[ht!]
\epsscale{1.0}
\plotone{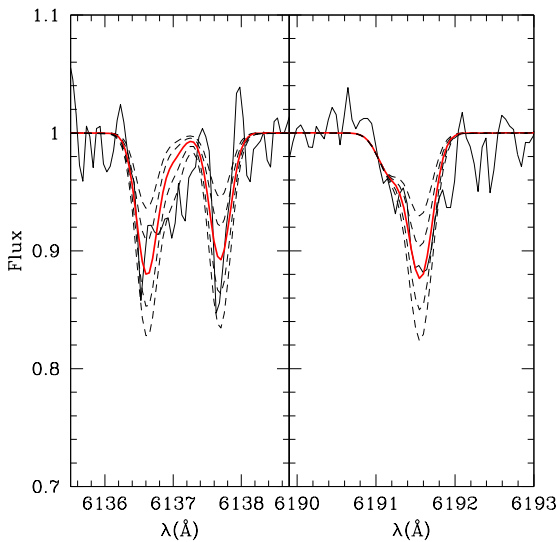}
\caption{Comparison between the spectrum of the star SGB$_{B}$.12 ([Fe/H]=-1.80) and 5 synthetic spectra
  having [Fe/H]=-2.30,-2.05,-1.80,-1.55, and -1.30. The best fitting
  spectrum is indicated with a thick red line.The star has
  F435W=18.4 and S/N=35, typical of the sample.}
\label{f2b}
\end{figure}

It has been suggested that this unusual He-rich population might come
from material contaminated by the ejecta of massive ($25 M_{\odot}$,
\citealt{no04}), or slightly less massive (10--14 $M_{\odot}$, \citealt{pi05})
supernovae, or from rapidly rotating low-metallicity massive stars
\citep{ma06}, or from intermediate-mass
asymptotic-giant-branch stars \citep{iz04}.

This double MS was not totally unexpected because \citet{no96} found a
bimodal distribution of [Ca/H] for RGB stars based on low-resolution
spectra, with a first peak at [Ca/H]$\sim$-1.4 and a second peak at
[Ca/H]$\sim$-1.0.
This result was partially confirmed in the same period by \citet{su96} that,
using the Calcium triplet method, found a [Fe/H] distribution with a peak at
[Fe/H]$\sim$-1.7 and a tail toward higher metallicities.

However Omega Centauri is much more complex than that,
because more than two stellar populations are present.
\citet{so05} could identify at least 4 stellar populations on the sub-giant branch (SGB)
having a mean [Fe/H]=-1.7,-1.3,-1.0, and -0.6 dex respectively, based on CaT abundances. \citet{vi07}
identified photometrically at least five
stellar populations in the SGB region, and spectroscopically three
populations based on their iron content, at [Fe/H]=-1.68, -1.37, and -1.14.

Several other studies tried to establish the number and iron content of the
populations.
The first was \citet{ca09}, based on Str\"{o}mgren photometry of the red-giant
branch (RGB). The authors found 6 peaks in the iron distribution at
[Fe/H]=-1.73,-1.29,-1.05,-0.80,-0.42, and -0.07 dex.
On the other hand \citet{jo10}, based on a metallicity distribution
obtained by a large number of high resolution RGB
spectra, identified four groups at [Fe/H]=-1.75,-1.50,-1.10,-0.75 dex.
Recently \citet{pa10} suggested also the presence of a very metal poor population,
at [Fe/H]=-1.95, based on high-resolution spectra.

Finally \citet{ma11} found that their metallicity distribution is consistent 
with the presence of multiple peaks corresponding to [Fe/H]=-1.75,-1.60,
-1.45,-1.00, a broad distribution of stars extending between -1.40 and -1.00,
and a tail of metal-rich stars reaching values of [Fe/H]-0.70.

In addition \citet{st06} found that an age range of 2-4 Gyrs exists in
the cluster, based on the position and metallicity of stars in the TO-SGB
region. This result was confirmed by
\citet{vi07}, who also suggested that a large age-spread
could affect the cluster. In particular they found that stars that belong to
the most metal-poor group ([Fe/H]$\sim$-1.7) span an age-range of several Gyrs
and that surprisingly the most metal-rich component is also the oldest.

After this brief summary it is clear that
a more complete study is required in order to better determine 
the number and mean metallicity of sub-populations in
$\omega$~Cen and its age-metallicity relation.
This is the best way to understand the complex
star-formation history of this intriguing object.
Such a study must be cafeful to try and account for any possible spread in He and CNO that affect
the stars of the cluster, as found \citet{no04} and \citet{ma11}.

The best region in the CMD for this purpose is the SGB, where the position
of a star strongly depends not only on the metallicity, but also on the age.
In this way both age and metallicity can be used to disentangle
and identify the sub-populations as well as to study any possible age-spread
and age-metallicity relation affecting them.

For this purpose we collected a large
spectroscopic database that covers the entire SGB of the central HST
photometric field (see \citealt{vi07}).
 
In Section~2 we present the observations and data reduction. 
In Section~3  we discuss the abundance measurements, while
Section~4 presents the results.
In Sections~5 our findings are compared with the results from the literature.
Finally Section~6 discusses the implications of the
observational facts presented in this paper for the stellar populations
in $\omega$~Centauri, the age spread, the age-metallicity relation, and the
origin of this anomalous cluster.

\section{Observations and data reduction}

The spectroscopic data come from the ESO proposal 082.D-0424(A),
and were collected in January--March 2009 with FLAMES@VLT\-+GIRAFFE. The sky
was clear, and the typical seeing was $\sim$0.8 arcsec (FWHM).  We used the MEDUSA
mode, which obtains 132 spectra simultaneously.  To have enough
$S/N$ and spectral resolution we used the HR13 set-up, which gives $R=22500$ in
the  6120--6405 \AA \ range.

\begin{figure*}[ht!]
\epsscale{1.7}
\plotone{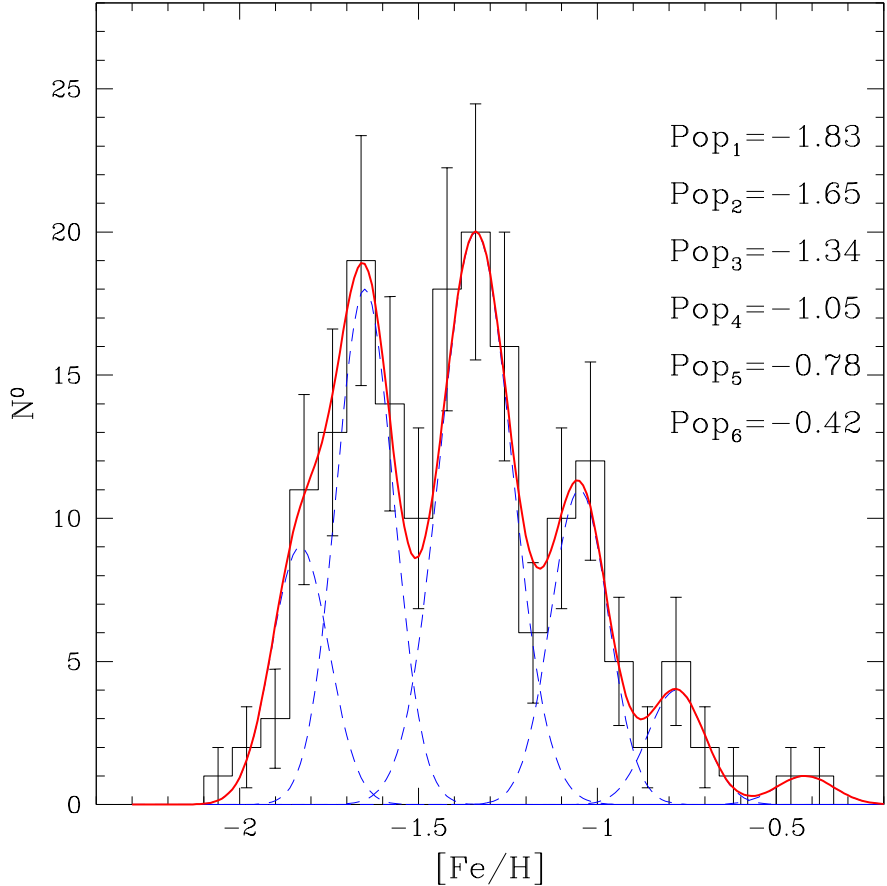}
\caption{[Fe/H] distribution of our entire sample.
  Each sub-population is represented by a blue Gaussian (dashed line)
  and its mean metallicity is indicated. The continuous red line is the
  sum of the Gaussians fitting the observational data.}
\label{f3}
\end{figure*}

The main target was the SGB, where we pointed 450 stars divided in 7
placements and observed 4 or 5 hours each.
The remaining fibers were placed on HB ($\sim$270) and RGB ($\sim$80) stars
and on the sky (10 fibers for each plate).
Results for the RGB targets were already presented in \citet{ma11} and \citet{ma12}.

SGB target stars were selected from the 3$\times$3 mosaic of {\sl
HST} fields presented in \citet{vi07} in order to cover the 5 SGBs
of the  $m_{\rm F435W}$,$m_{\rm F625W}$ CMD identified in that paper.
In this paper we call the 5 SGBs as: A,B,C,D, and E (see Fig.~\ref{f1}).

Stars were selected for the observations in order to have no
neighbors closer than 0.6 arcsec and brighter than 
$m_{\rm F625W}$+2.5 mag, where $m_{\rm F625W}$ is the
magnitude of the target, to avoid possible contamination. 
Target stars were then further cleaned for any remaining contamination
during the data analysis process as explained in the following section.
We identified our targets also in the ground-based photometry by \citet{be09}.
This was done to obtain V magnitudes, needed to estimate gravity (see next
section). However, in a few cases we found that the V ground based magnitude
was widely discrepant compared to $m_{\rm F625W}$, probably because of
the crowding.
In order to remove this problem we decided to used an {\it interpolated} 
V magnitude (V$_i$). To do that we plotted V vs $m_{\rm F625W}$. The relation
is linear, so we fitted a straight line using a 3$\sigma$ clipping rejection
algorithm. Finally we adopted V$_i$ obtained from this relation and the appropriate
$m_{\rm F625W}$ magnitude of each star.

\begin{figure*}[ht!]
\epsscale{1.7}
\plotone{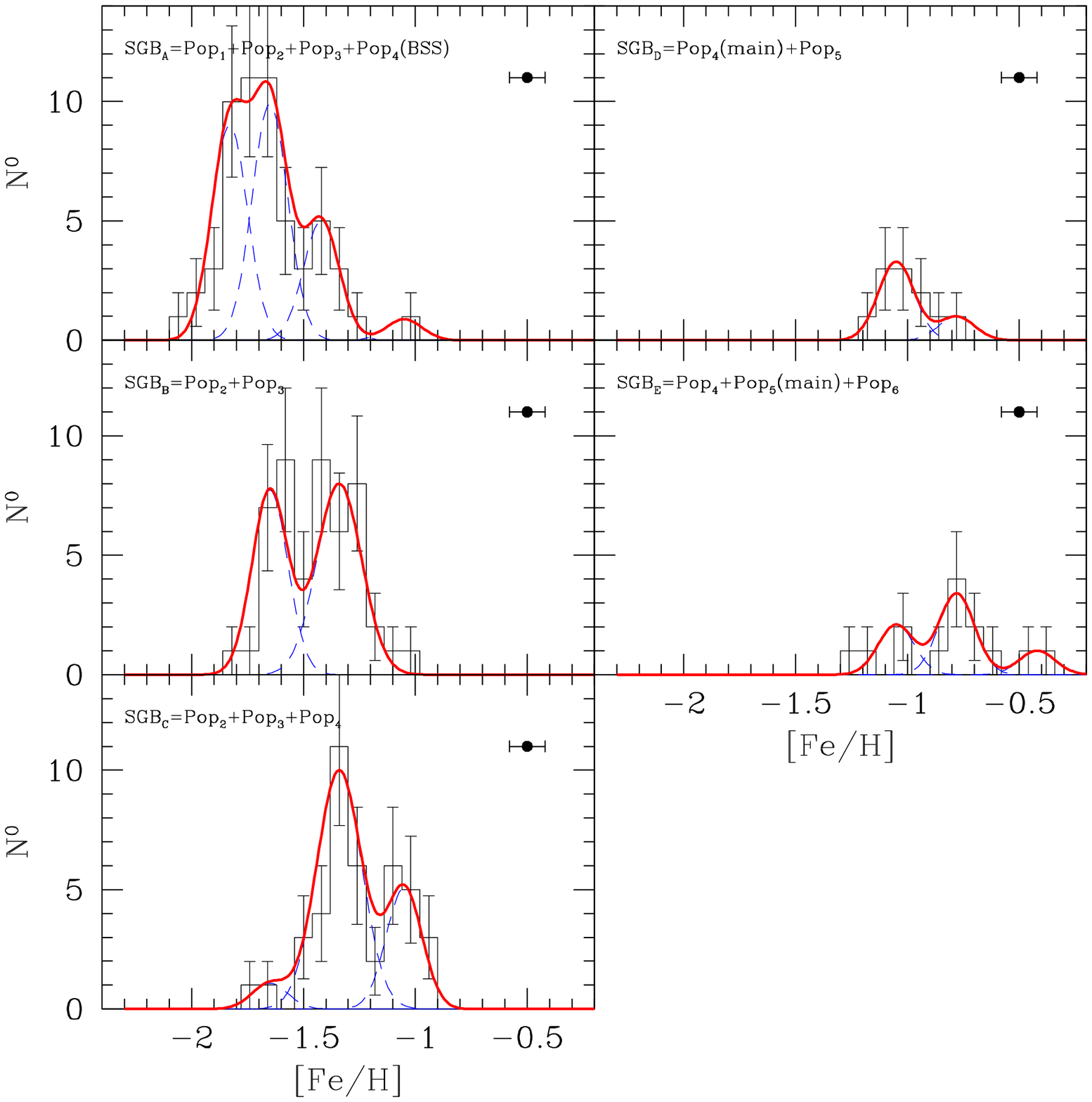}
\caption{[Fe/H] distribution of each SGB.
  Each sub-population is represented by a blue Gaussian (dashed line).
  The continuous red line is the sum of the Gaussians fitting the observational
  data. For each SGB we indicated the populations that we could fit. Errorbars
  in the top-right part of each panel represent the observational error on the [Fe/H] measurement.}
\label{f4}
\end{figure*}

The data were reduced using GIRAFFE pipeline 1.13 \citep{bl00},
which corrects the spectra for bias and flat-field.  (See
http://girbldrs.sour\-ceforge.net/ for documentation on the GIRAFFE
pipeline and software.).  A sky correction was applied to each stellar spectrum by
subtracting the average of ten sky spectra that were observed
simultaneously with the stars (same FLAMES plate).  The wavelength
calibration uses calibration-lamp spectra taken the following day with respect
to the observations.
Finally, each spectrum was normalized to the continuum.
The resulting spectra have a dispersion of 0.05 \AA/pixel and a typical
$S/N\sim$25-40, with a median value of 35.

We used the {\sf fxcor} utility of IRAF to
measure the radial velocity, which we then converted to heliocentric.
The error in radial velocity is typically about $\sim$1 km/s.
Considering the mean radial velocity of $\omega$~Cen ($\sim$232 km/s, \citealt{re06}), the
velocity dispersion in the inner part of the cluster ($\sim$15 km/s,
\citealt{re06}), and the observational errors, all of the stars
with radial velocity in the range 180-300 km/s were considered members.

The coordinates, magnitudes, and radial velocities of our members 
are reported in Tab~\ref{t2}. This table reports on the final targets after eliminating 
binaries and contaminated objects, as explained in the following section.

\begin{figure*}[ht!]
\epsscale{2.0}
\plotone{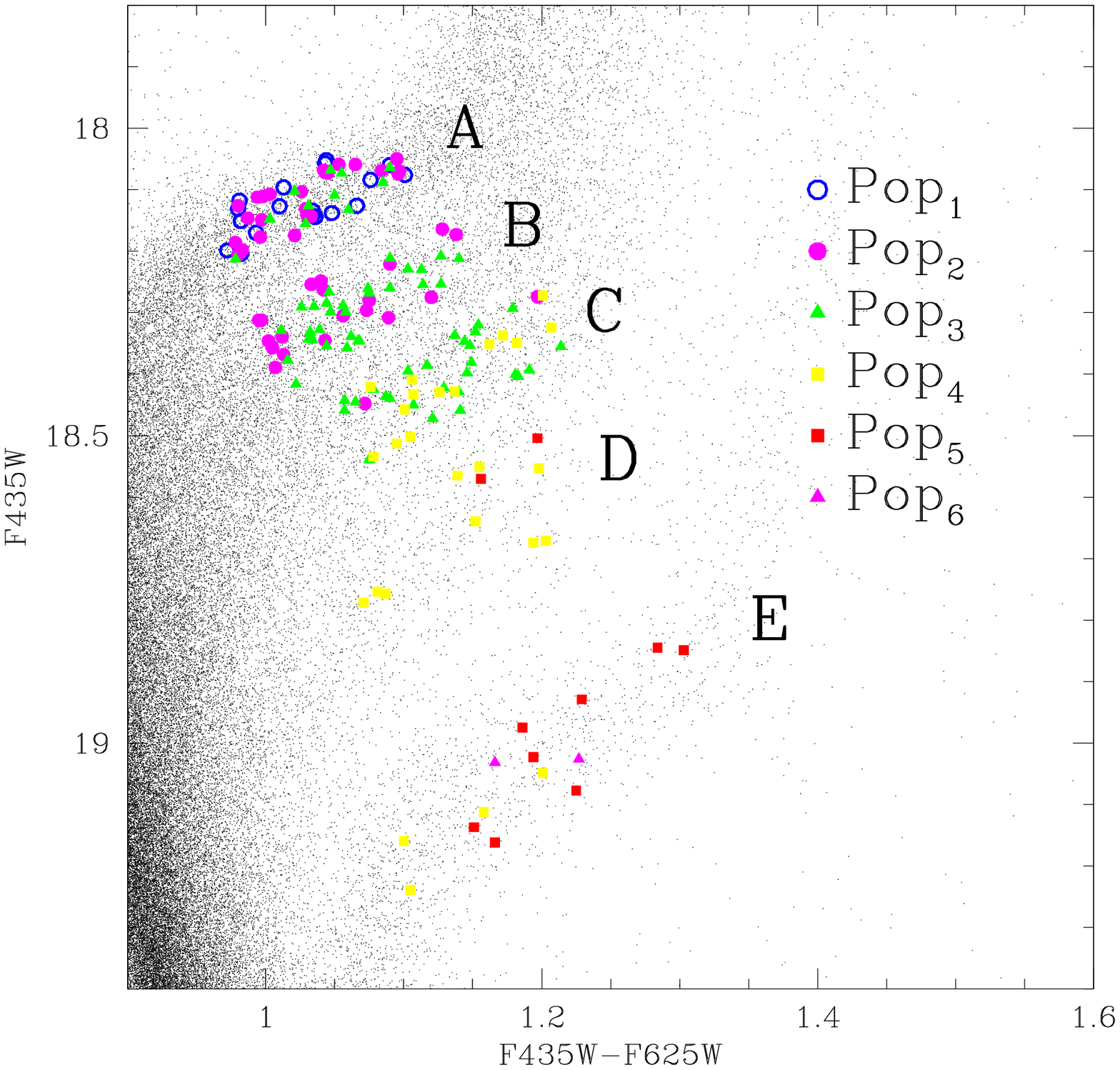}
\caption{Position on the CMD of the subpopulations identified in this paper.}
\label{f10}
\end{figure*}

\section{Abundance measurements}

First of all we checked for any possible residual contamination of our
targets by neighbor stars that can be easily identified in our photometry. 
To be conservative and avoid any misinterpretation of the data, we finally decided to
retain only those targets with no contamination, i.e. those targets
with no neighbors brighter than $m_{\rm F625W}$+2.5 mag within 3 times the
FWHM of our observations. This guarantees us that the final metallicity is not
altered by contamination effects. We had to reject 264 stars.

Second, we checked for possible binaries looking for radial velocity
variations. We expect most binaries to be composed of a SGB and a MS; such
systems should be brighter on average
with respect to the single star sequence. For this reason they would appear
younger than they really are. For each target we have 4/5 radial velocities
obtained in different epochs, with an epoch range of a few weeks. So we obtained the r.m.s. for each star and
finally the r.m.s. distribution (not reported here). According to this
distribution we flagged as binaries all those stars that show a r.m.s. larger
than 7 km/s. We rejected 14 stars as binaries.

After the contamination and the binary checks, we were left with 172 objects 
that are plotted in Fig.~\ref{f1}.

As in \citet{vi07}, we derived effective
temperatures ($T_{\rm{eff}}$) from the $m_{\rm F435W}$ $-$ $m_{\rm
F625W}$ color in the {\sl HST} CMD.  The relation between color and
$T_{\rm{eff}}$, as a function of [M/H] (by which we mean the global
metallicity, including alpha elements), was derived from isochrones
by \citet{pi06}\footnote{http://basti.oa-teramo.inaf.it/index.html}.  Colors were de-reddened using the
absorption coefficients listed in Table~3 of \citet{be05},
adopting $E(B-V)=0.115$.  As a first guess for the [M/H] to be used in
the color-[M/H]-temperature relation, we adopted [Fe/H] = $-$1.5, the
mean metallicity of $\omega$~Cen stars, along with an $\alpha$-enhancement
of 0.4 dex for all stars (see \citealt{jo10}). 
The [M/H] was derived from the adopted [Fe/H] and
the alpha enhancement from the prescription by \citet{sa93},
along with the corresponding $T_{\rm{eff}}$ from the
color-[M/H]-temperature relation.  Using this value for $T_{\rm{eff}}$,
we calculated $\log g$ and $v_{\rm{t}}$ and measured a new [Fe/H]
abundance as described below.  Then for each star the values of
$T_{\rm{eff}}$ and [Fe/H] were changed in an iterative process, till
convergence (when $\log g$ and $v_{\rm{t}}$ change less than 0.02 dex and 0.02
km/s respectively).

As noted by \citet{vi07}, the effect of variations in helium content on the
relation between color and temperature is of the order of
$\sim$10 K in temperature for SGB stars, which translates into a change
of $\sim$0.01 dex in metallicity. Such small changes can be neglected.

The gravity $\log g$ was calculated from the elementary formula
$$ \log\left(\frac{g}{g_{\odot}}\right) =
         \log\left(\frac{M}{M_{\odot}}\right)
         + 4 \log\left(\frac{T_{\rm{eff}}}{T_{\odot}}\right)
         - \log\left(\frac{L}{L_{\odot}}\right). $$
The mass $M/M_{\odot}$ was derived from isochrone fitting.
The luminosity $L/L_{\odot}$ was derived
from V$_i$, assuming the absolute distance modulus $(m-M)_0=13.75$ found by \citet{va06},
and the reddening adopted above.  The bolometric
correction (BC) was derived from the BC-$T_{\rm{eff}}$ relation of
\citet{al99}.  Finally, the microturbulence velocity came from
the relation \citep{gr96}:
$$v_{\rm t} = 2.22 - 0.322 \log g.$$

The adopted atmospheric $T_{\rm{eff}}$, $\log g$, and v$_t$ are listed in Tab~\ref{t2}.

The metal content was obtained by comparison with synthetic spectra
calculated using MOOG \citep{sn73}. 
The model atmospheres of Kurucz (1992), used throughout this paper, assume
$N_{\rm{He}}/N_{\rm{H}} = 0.1$, corresponding to $Y=0.28$ by mass.  The
bMS and the related SGB stars (i.e., stars with the
same metallicity as the bMS stars), are thought to have a helium
content $Y\sim0.38$  As discussed in \citet{pi05}, the variation in 
the atmospheric structure due to this increase in helium introduces an
systematic error smaller than 0.03 dex in the metal-abundance 
determinations, which is negligible.  

Our [Fe/H] values were obtained from a comparison of each observed
spectrum with five synthetic ones (see Fig.~\ref{f2b}), calculated with
different metal abundances. We used the regions at 6136-6138 \AA\ and at
6191 \AA\ for this purpose. These regions contain the only Fe lines not
contaminated by telluric absorption and emission features and visible in all
our spectra due to their S/N and to the T$_{\rm eff}$ and low metallicity of the
targets. In the most metal-rich spectra other iron lines are visible, but to
be homogeneous we had to choose those that are the only visible also in the
most metal-poor targets. 
log(gf) of the lines were calibrated by spectrosynthesis on the
Sun assuming log$\epsilon$(Fe)=7.50 and the solar spectrum by \citet{ku84}.

The metallicity was obtained by  minimizing the
r.m.s.\ scatter of the differences between the observed and synthetic
spectra. [Fe/H] values derived for our targets are listed in Tab~\ref{t2}.

At this point we must note the the [Fe/H] we obtained is based on the
assumption of a standard $N_{\rm{He}}/N_{\rm{H}}$ content. However the cluster has a mean He
content that varies from Y=0.25 for the most metal poor stars, to Y$\sim$0.39
for the most metal rich, according to \citet{jo13}.
A larger He content implies a lower H content, and by consequence the real
[Fe/H] value should be higher than that we obtained. For $\Delta$Y=0.14 we
should apply a correction of $\Delta$[Fe/H]={\bf +}0.09. However this would make the
comparison with the literature difficult, so we prefer to use our present value
and just warn the reader about this effect.

Because we have uncontaminated objects,
the random internal error is dominated by the noise of the
spectra. S/N ranges from 30 to 40 for most of the stars, while the faintest
have S/N$\sim$25 in the worst case.
To estimate the error in [Fe/H] we used Monte Carlo simulations.
For this purpose we calculated a spectrum representative of the two most
populous and extreme populations, the brightest at [Fe/H]$\sim$-1.8, 
and the faintest at [Fe/H]$\sim$-0.8.
Then we added noise to each one in order to obtain
1000 new spectra that simulate the real ones. Finally we measured the [Fe/H]
of each simulated spectrum. Both for brighter and fainter targets we
found $\sigma$([Fe/H])=0.08 dex. This is probably because the decrease of S/N is
compensated by the increase of the mean metallicity, which gives stronger
spectral lines.
Another way to estimate this error is to use the formula by \citet{Ca88}:

$$ \sigma_{\rm EQW} \sim 1.06 \times \sqrt(FWHM\cdot \delta x)/(S/N) $$

In our case FWHM=0.35\ \AA and $\delta$x=0.05 \AA. For the median S/N of our
spectra ($\sim$35), the expected $\sigma_{\rm EQW}$ is 4 m\AA. This translates
into $\sigma$([Fe/H])=0.07 if we consider that we used three iron lines to
estimate abundances, which is close to the value we obtained with the Monte
Carlo simulations.
 
To this error we should add
(in quadrature) the error due to photometric uncertainty in the colors;
the error in color is typically of the order of 0.01 magnitude, which
translates into a 0.02 dex error in abundance (see \citealt{vi07}). Finally we adopt an overall
uncertainty of 0.08 dex for [Fe/H].  This is the
internal random error in our metallicity measurement. In addition there can be
a systematic error of the order of 0.15--0.20 dex, because of systematic
uncertainties in the effective temperature scale, in the model atmospheres,
in the distance and reddening. The systematic errors do not affect the
relative metallicities of the different stellar populations of
$\omega$~Cen that we will discuss in later sections.

\section{Results}

First of all we plot the iron distribution of the entire
sample in Figure~\ref{f3}. Then in the following plot 
(Fig.~\ref{f4}) we report the iron distribution
of the five SGB branches of figure~\ref{f1} separately.

The first basic thing we note is that according to our analysis any of the SGBs
has a distribution in metallicity with a spread that exceeds the observational
errors. Such an error is plotted in Fig.~\ref{f4} as an errorbar in the
top-right part of each panel.
In addition each SGB displays several peaks, that allowed us to
identify a certain number of subpopulations that form the cluster.

The identification of the metallicity-based sub-populations was done
in order to reproduce at the same time the metallicity distribution of
the entire sample and the metallicity distribution of the single SGBs.
We could identify 6 of them termed:

$$Pop_1: [Fe/H]=-1.83$$
$$Pop_2: [Fe/H]=-1.65$$
$$Pop_3: [Fe/H]=-1.34$$
$$Pop_4: [Fe/H]=-1.05$$
$$Pop_5: [Fe/H]=-0.78$$
$$Pop_6: [Fe/H]=-0.42$$

Their presence and their mean [Fe/H] value are justified by the following analysis.

In the figures each subpopulation was fitted with a Gaussian (blue dashed line)
having  a $\sigma$ that was allowed to vary up to $\pm$0.02 dex around our
theoretical uncertainty of 0.08 dex in order to obtain a better fit. 
In each figure the continuous red line is the sum of all the
single Gaussians. We allowed also the mean [Fe/H] value of each
subpopulation to vary of few hundredth of dex 
in order to obtain the best match possible with the data. 

Fig.~\ref{f4} shows that
SGB$_{A}$ is dominated by Pop$_1$ and Pop$_2$, but Pop$_3$ is clearly visible. 
In addition there is a faint peak that could correspond to Pop$_4$.
The most probable explanation is that this peak corresponds to evolved blue
straggler stars (BSS).

SGB$_{B}$ is composed of a mix of Pop$_2$ and Pop$_3$.

In SGB$_{C}$ Pop$_3$ and Pop$_4$ are clearly visible 
as two well defined peaks. The histogram shows also a
tail with a secondary peak at [Fe/H]$\sim$-1.7. 
We interpret this as Pop$_2$.

SGB$_{D}$ is dominated by Pop$_4$, but there is a tail that corresponds
to Pop$_5$. 

SGB$_{E}$ is dominated by Pop$_5$, but Pop$_4$ is clearly visible as a
well defined tail, while Pop$_6$, in spite of being formed by only
2 stars, forms a separated peak. 

If we consider Fig.~\ref{f3}, we can clearly identify Pop$_2$, Pop$_3$,
Pop$_4$, Pop$_5$, and Pop$_6$ as well defined peaks. Pop$_1$ forms a 
low metallicity tail and not a peak, but if we remove it we cannot fit
properly the iron distribution.

\section{Comparison with the literature}

We already presented in the introduction the most recent papers that
discuss the number and iron content of the sub-populations of $\omega$ Cen.
In this section we discuss how those results can be interpreted in the light
of what we have found here. We underline the fact that, due to the very
extensive literature and to the very different methodologies used to study this
cluster, we focus our attention only on those papers that try to identify the
number of sub-populations based on their [Fe/H] content.

\citet{so05} found  4 stellar populations at [Fe/H]=-1.7,-1.3,-1.0, and -0.6.
Looking at their figure 4, we can suggest that their population at -1.7 is
a mixture of Pop$_1$, and Pop$_2$, that at -1.3 is our 
Pop$_3$, that at -1.00 is our Pop$_4$ and that at -0.6 is our Pop$_5$.

\citet{vi07} found 3 stellar populations at [Fe/H]=-1.68,-1.37, and -1.14.
Looking at their figure 15, the population at -1.68 can be identified
as a mixture of Pop$_1$ and Pop$_2$, while that at -1.37 is our
Pop$_3$. The group at -1.14 is our Pop$_4$.

\citet{ca09} found 6 peaks in the iron distribution at
[Fe/H]=-1.73,-1.29,-1.05,-0.80,-0.42, and -0.07 dex.
Looking at their figure 17, the group at -1.73 can be identified
as a mixture of Pop$_1$, and Pop$_2$. In particular Pop$_1$
is visible as a peak at [Fe/H]$\sim$-1.8$\div$-1.9, not pointed out by the
authors. The peak at -1.29 is our Pop$_3$, while the peak at -1.05 is our
Pop$_4$. Their peaks at -0.80 and -0.42 are our Pop$_5$
and Pop$_6$ respectively. In particular the identification of the same
population at [Fe/H]=-0.42 in two independent datasets
make us confident that Pop$_6$ is real.
On the other hand we do not have any trace of their peak at -0.07.
An explanation could be that the corresponding population forms a weak SGB branch fainter than SGB$_E$,
so we did not recognize it in our CMD, and we did not point any fiber on
its stars, or that it is too centrally concentrated, so we missed it in the
fiber pointing

\citet{jo10} identified four groups at [Fe/H]=-1.75,-1.50,-1.10,-0.75.
Looking at their figure 8, the populations at -1.75, -1.10, and -0.75 can be 
identified with our  Pop$_1$+Pop$_2$,  Pop$_4$, and  Pop$_5$ respectively.
Their peak at -1.50 does not correspond to any of our populations.

We do not confirm the presence of a very metal poor population ([Fe/H]$\sim$-1.9), as
suggested by \citet{pa10}. However our data do not have the required accuracy
and statistics to identify such a feature.

Finally we compare our results with \citet{ma11}.
These data show clear peaks at [Fe/H]=-1.76, -1.60,-1.00, and -0.76 that
correspond to our Pop$_1$,Pop$_2$,Pop$_4$, and Pop$_5$ respectively.
Pop$_3$ is visible as a tail of Pop$_2$.

\begin{deluxetable}{lccc}
\tablecolumns{12}
\tablewidth{0pc}
\tablecaption{Most probable [Fe/H] range, SGB correspondence, and age-spread for each sub-population.}
\tablehead{
\colhead{Population} & \colhead{[Fe/H]} & SGB & \colhead{Age-spread (Gyrs)} 
}
\startdata
\hline
Pop$_1$ & -1.83 &  A       & 0\\
Pop$_2$ & -1.65 &  A+B(+C) & 2(4)\\
Pop$_3$ & -1.34 &  A+B+C   & 4\\
Pop$_4$ & -1.05 &  C+D+E   & 7\\
Pop$_5$ & -0.78 &  E(+D)   & 0(4)\\
Pop$_6$ & -0.42 &  E       & 0\\
\hline
\enddata
\label{t1}
\end{deluxetable}

\section{The age spread and the age-metallicity relation}

After having determined the sub-populations that form the
cluster and their distribution on the SGB, we can discuss 
the implication of the present results on the age spread 
that, according to \citet{vi07}, affects the cluster and that should
be of the order of several Gyrs.
This is a controversial topic, because some authors suggest
that $\omega$ Cen does not have any age spread at all (e.g. \citealt{so05}).

We start by estimating the range in magnitude each sub-population covers
at the level of the SGB. We do not consider peaks identified as BSS.

In order to visualize the spread in magnitude of each
population, we plot in Fig.~\ref{f10} their position on the SGB.
The membership of each star was decided based on which Gaussian dominates 
at its metallicity in Fig.~\ref{f4}. 
The [Fe/H] interval assigned to each population changes from one SGB
branch to the other in order to minimize the contamination due to measurement errors.
Of course some contamination remains, but with the given error in
metallicity it is impossible to separate completely the six groups of stars.
For the following discussion we assume the each population has no intrinsic
[Fe/H] spread, and that the enlargement associated to each peak in the 
[Fe/H] distribution histograms is totally due to the measurement error.
This is justified by the fact that Gaussians with a
$\sigma$ of 0.08 dex (that is our internal measurement error) well fit the
total [Fe/H] distribution of Fig.~\ref{f3}. 

Pop$_1$ forms only SGB$_A$.
Pop$_2$ forms part of SGB$_A$, SGB$_B$, and maybe SGB$_C$, so its
spread is of the order of 0.2$\div$0.4 mag. 
Pop$_3$ forms part of SGB$_A$, SGB$_B$, 
and SGB$_C$ so its spread is of the order of 0.4 mag. 
Pop$_4$ forms SGB$_C$, SGB$_D$ and SGB$_E$ so its spread is of the order of
0.7 mag.
Pop$_5$ forms SGB$_E$ and maybe SGB$_D$ so its spread is $<$ 0.4 mag.
Pop$_6$ does not show any spread and all its stars belong to SGB$_E$.

The correspondence between populations and the five SGBs is given in
Tab.~\ref{t1}.

Before giving an estimation of the age spread of each population we must
address some further considerations. 

The first concerns the interval in age that corresponds to a $\Delta$$m_{\rm F435W}$ 
of 0.1 mag on the SGB. In \citet{vi07} we already performed this exercise, and it turns out
that the exact value depends both on metallicity and helium content. However
a good aproximation is $\sim$1 Gyr per 0.1 mag. 

The second concerns the total CNO content of each population. NGC1851 has
a double SGB \citep{mi08}, where the two population are separated of about 0.1
mag in luminosity. This implies an age difference of about 1 Gyr if the two
populations have the same CNO. However, \citet{ca08} showed that a CNO
difference of 0.3 dex can reduce the age difference to 100 Myrs.

\begin{figure}[ht!]
\epsscale{1.0}
\plotone{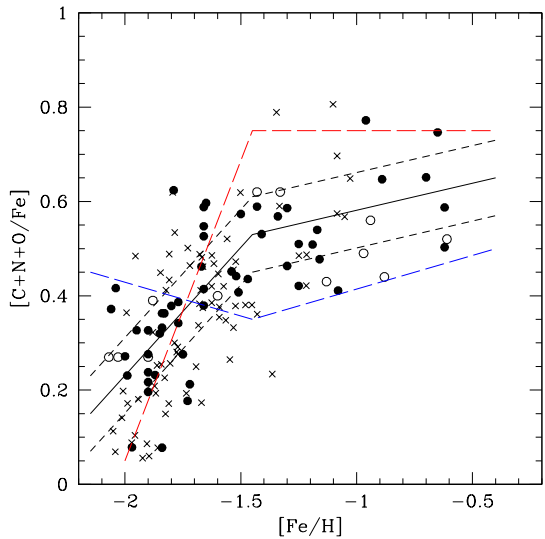}
\caption{Mean trend of [C+N+O/Fe] vs. [Fe/H] for stars in $\omega$ Centauri.
See text for more details.}
\label{f11}
\end{figure}

To investigate this point we plotted in Fig.~\ref{f11} the total CNO content as
a function of the metallicity ([Fe/H]) using data from \citet[crosses]{ma12},
\citet[open circles]{do11}, and Villanova et al. (2014, in preparation, filled points).
\citet{ma12} and Villanova et al. (2014) well sample the region
below [Fe/H]=-1.0, while \citet{do11} and Villanova et al. (2014) are
used to sample the more metal rich part above [Fe/H]=-1.0.
We draw the mean trend that is represented by the black continuous
line, while the black dashed lines are the mean trend shifted vertically by $\pm$0.08
dex, that is the mean r.m.s. of the data (see below). 
We notice that the mean CNO content has two different linear trends.
the first, from [Fe/H]$\sim$-2.0 to [Fe/H]$\sim$-1.5 has a slope of
$\sim$0.76, while the other from [Fe/H]$\sim$-1.5 to [Fe/H]$\sim$-0.5 is
flatter with a slope of $\sim$0.12. 
After that we calculate the spread of the points around the mean trend, and
found that the dispersion of [CNO/Fe] is $\sigma$=0.08 dex. This is a
small value, comparable with the typical measurement error of any chemical
abundance determination procedure (e.g. \citealt{vi11}). So we can assume that 
any intrinsic spread in CNO (at a given metallicity), if present at all, 
is negligible. A more straightforward procedure would be the direct
calculation of the measurement error on the [C+N+O/Fe] quantity, that depend
on the error on temperature, gravity, metallicity, microturbulence, and on the
S/N, but this is out of the aim of this paper.
Our conclusion is that we can assume a constant CNO content within each
sub-population.

We can now translate the magnitude spread 
into an age spread. Results are reported in Tab.~\ref{t1}.
Three populations (Pop$_2$, Pop$_3$, and Pop$_4$) have an age-spread of at least 2 Gyrs.
Pop$_4$ shows a surprisingly large age-spread.

Pop$_1$ and Pop$_6$ do not show an age-spread. On the other hand Pop$_5$
possibly occupies also SGB$_D$, but the number of its possible stars is so
small (a total of 2) that we prefer to live the question open and assign it an age-spread of 0,
and to put the possible membership to SGB$_D$ and the correspondent age-spread within
parenthesis in Tab.~\ref{t1}. Also Pop$_2$ suffers the same problem and its
membership to SGB$_C$ is uncertain. So we decided again to put it within parenthesis
in Tab.~\ref{t1}, together with the relative age-spread.

So we conclude that $\omega$ Centauri shows clear evidence of a
significant age-spread, at least for three of its populations, when they are
identified based on metallicities alone.

The next step is to transform all the information we have, i.e. the
metallicity and magnitude of each star, in an age-metallicity relation,
taking advantage of the fact that the SGB is the place in the CMD most
sensitive to age effects. The aim is to transform the F435W magnitude of each
star into an age. However there are several effects that can be taken into
account using isochrones \citep{pi06}. The
most obvious one is the fact that stars of the same age but different
metallicity have different F435W magnitude, with the most metal rich being
also the faintest. We found
that a difference of 1.0 dex in metallicity corresponds to a F435W difference of
0.64 mag on average, taking the other parameters constant. 
On the other hand the $\alpha$-enhancement is not an issue because
all stars have the same $\alpha$ content at all metallicities \citep[Fig. 13]{jo10}.
A further effect to take into account is the He content.
In \citet{pi05} we showed that stars with [Fe/H]$\sim$-1.7 have a normal He
content (Y$\sim$0.25), while stars with [Fe/H]$\sim$-1.4 are He-enhanced
(Y$\sim$0.38). Recently \citet{jo13} published a more detailed He trend that
we adopted here. According to this trend, all stars with [Fe/H]$<$-1.55 have Y=0.25, all stars
with [Fe/H]$>$-1.31 have Y=0.39, while for stars in between Y linearly
increases from [Fe/H]$\sim$-1.55 to [Fe/H]$\sim$-1.31. Again, using isochrones
with different He content, we found that $\Delta$Y=+0.1 corresponds to an
increase of F435W of $\sim$0.05 mag, taking the other parameters constant. We notice that the large uncertainty on the
He-[Fe/H] relation is compensated by the small effect of Y on the F435W
magnitude (much lower than that due to the metallicity).
The CNO trend was already discussed above. We just take it into account using the
relation published by \citet[section 5]{ma12} that is valid for [Fe/H]$<$-1. In this iron
regime the effect of CNO enhancement is independent of metallicity. For
[Fe/H]$\sim$-0.4 (the upper limits of our metallicity range) isochrones show
that the effect of CNO enhancement on age is 3 times larger than for the
[Fe/H]$<$-1 regime, taking the other parameters constant. We linearly
interpolated in between.

\begin{figure*}[ht!]
\epsscale{2.0}
\plotone{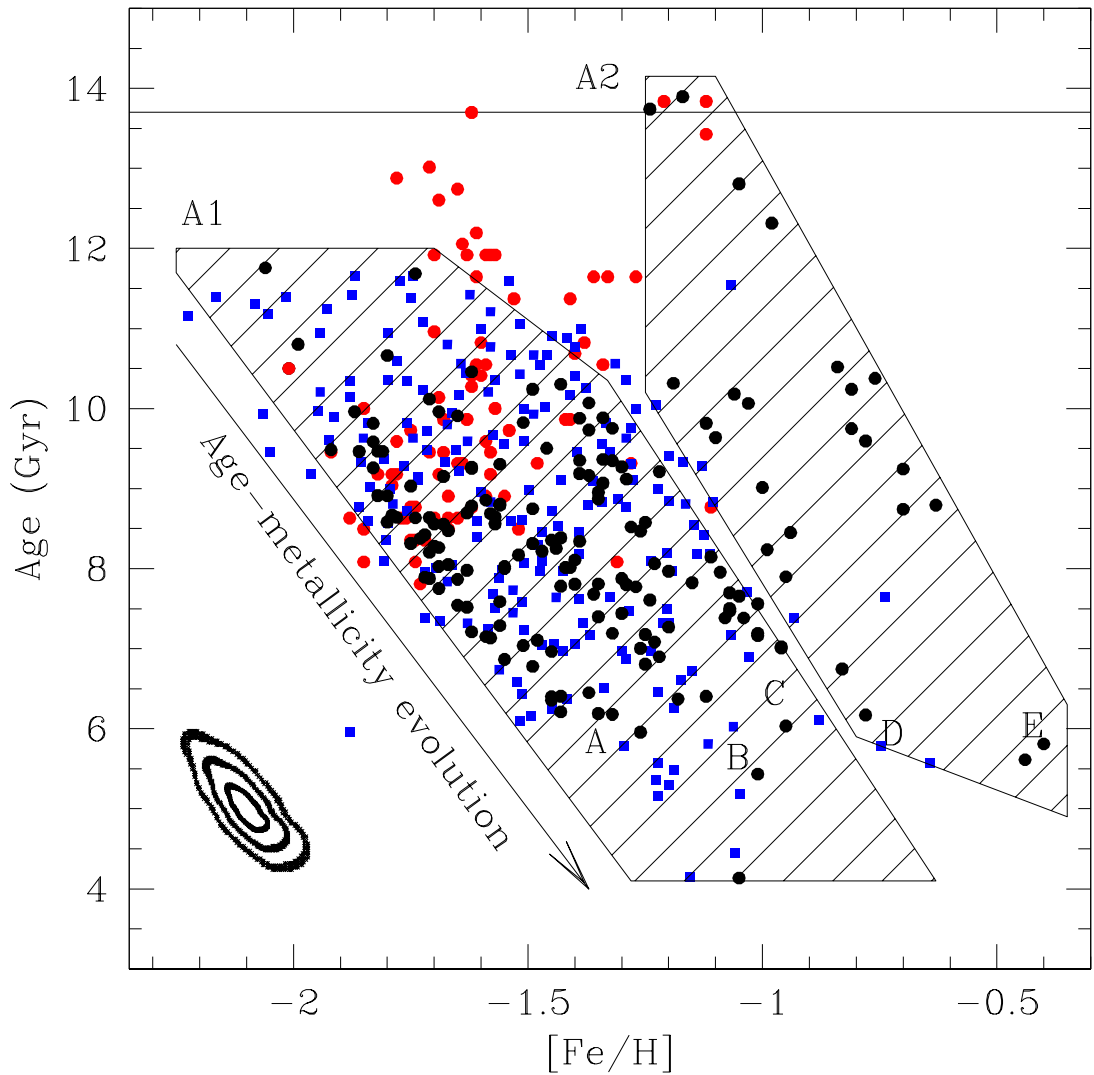}
\caption{The age-metallicity relation for $\omega$ Centauri. 
Contours represents the 1$\sigma$, 2$\sigma$, and 3$\sigma$ errors
respectively. Red points are data from \citet{vi07}, while blue points are
data from \citet{hi04}. See text for more details.}
\label{f12}
\end{figure*}

Finally, as already said, for a given metallicity, He content, and C+N+O
content, a 0.1 mag. difference in F435W corresponds to an age difference of
$\sim$1 Gyr, taking the other parameters constant. We verified also that
all the quantities we assumed to transform
F435W into age apart from the CNO content (i.e. 0.64 mag/dex for the metallicity,
0.05 mag for a change of 0.1 in Y) are fairly constant over the entire
interval of ages and metallicities covered by our stars. They are reported in
Tab.~\ref{t3} for reference. 

\begin{deluxetable}{llc}
\tablecolumns{12}
\tablewidth{0pc}
\tablecaption{The sensitivity of F435W magnitude on metallicity, Y,and age
(first three raws), and the sensitivity of age on CNO content (last two raws).}
\tablehead{
\colhead{Parameter} & \colhead{Variation} &  \colhead{F435W/Age variation} 
}
\startdata
\hline
$\Delta$[Fe/H]              & $\pm$1.0 dex & $\pm$0.64 mag\\
$\Delta$Y                   &    +0.10     & $\pm$0.05 mag\\
$\Delta$Age                 & $\pm$1.0 Gyr & $\pm$0.10 mag\\
$\Delta$CNO([Fe/H]$\leq$-1.0)    & $\pm$1.0 dex & $\mp$3.3 Gyr      \\
$\Delta$CNO([Fe/H]=-0.4)    & $\pm$1.0 dex & $\mp$9.9 Gyr      \\
\hline
\enddata
\label{t3}
\end{deluxetable}

In order to take into account also the SGB tilt (see Fig.~\ref{f1}), we proceeded as
in \citet{vi07}, i.e. we fitted a straight line to the upper SGB, and
calculated the distance of each star with respect to this line. We notice that
the five SGBs of Fig.~\ref{f1} are not perfectly parallel to each other, but
this does not affect our final result in a significant way.
All the quantities discussed before (i.e. 0.64 mag/dex,
0.05 mag for a change of 0.1 in Y, and 0.1 mag/Gyr) are related to a
vertical difference of the magnitude F435W in the CMD, and for this reason
they were transformed to this new reference system parallel to the upper SGB too.

Error on the final age is a function of the [Fe/H] difference between two
stars. For targets of the same metallicity and, as a consequence, of the same
He and CNO content, it is dominated by the error on the magnitude and [Fe/H]
value. To estimate the error in this case we used a Monte Carlo simulation.
We took an artificial star representative of the entire sample assigning it a
metallicity of [Fe/H]=-1.50 a magnitude $m_{\rm F435W}$=18.5 and a color 
$m_{\rm F435W}$ $-$ $m_{\rm F625W}$=1.12. After that we generated 10.000 stars
according to a random Gaussian distribution centered on these values and with a 
dispersion $\sigma$ on $m_{\rm F435W}$ and $m_{\rm F625W}$ of 0.01 mag (the
typical photometric error for a SGB star), and on
the metallicity of 0.08 dex. Finally we estimated the age of these 10.000 stars
using the same method described above for the real stars.
The result is the errors reported in Fig.~\ref{f12} as contours. The inner
contour is the 1$\sigma$ error, the contour in the middle is the 2$\sigma$ error,
while the outer contour is the 3$\sigma$ error.

For completeness we should add the random error due to some possible
He spread for a given Fe abundance that overlaps the He-Fe relation obtained 
by \citet{jo13}. Let's take an hypothetic star in the middle of the He range we assumed,
i.e. a star with Y=0.32, and consequently [Fe/H]=-1.43. A difference in its He 
abundance from the adopted value would not influence its age directly because
the dependence of age on He is very weak,
but would affect its age determination through its [Fe/H] value. In fact a
larger He content implies a lower H
content, and by consequence the real [Fe/H] value should be higher than that we
would have obtained using the method of section 3. A reasonable assumption is
that our hypothetical star can have an He value in the range of
$\Delta$Y=$\pm$0.07 around Y=0.32. This is half of the total He
interval obtained by \citet[$\Delta$Y=0.14]{jo13}. A fast calculation shows that this
corresponds to a [Fe/H] correction of $\pm$0.045 dex. This is negligible
compared to our error on [Fe/H] of 0.08 dex. It is also an overestimation
because we should use not $\Delta$Y=$\pm$0.07 for our purposes, but the
$\sigma$ of the He spread that, if the He distribution is a Gaussian, is of
about 0.14/6$\sim$0.02. This makes the impact of any possible He spread for a
given metallicity totally negligible compared with the error shown in Fig.~\ref{f12}.

For stars at the extremes of the [Fe/H] interval, uncertainties in the He and CNO 
trends also must be considered. instead. While the impact of the He trend uncertainty is
negligible, the impact of the CNO trend uncertainty will be discussed in the
next subsection.
 
We underline the fact that the ages obtained so far are relative ages, and
errors we estimated are errors on relative ages. Absolute ages were obtained
applying a rigid shift to the whole sample in order that the oldest stars have the
age of the Universe. Systematic error on absolute ages is surely larger than
1 Gyr, but for our purposes it is not a concern.

The final product of this procedure is the age-metallicity relation presented
in Fig.~\ref{f12}. Ages derived for our targets are listed in Tab~\ref{t2}. 
Our data (black points) appear to follow five well defined
strips. This is not a real effect but a consequence of our target selection
that was focused on the five SGBs. The correspondence between
the five SGBs and the five strips is indicated by the five letters that identify
the SGBs in Fig.~\ref{f1}. 
Red points are the age-metallicity relation by \citet{vi07}.
We apply a vertical shift to our points in order
to match the age of the three old metal rich stars ([Fe/H]$\sim$-1.15,
age$\sim$13.7 Gyrs). We note that red points in the -1.8$<$[Fe/H]$<$1.3
regime are older than black points. This is because in \citet{vi07} we did
not take into account the affect of the C+N+O content.

Blue points are the age-metallicity relation by
\citet{hi04}. Again, we shifted vertically these data in order to
match the black points. 

The easiest way to interpret Fig.~\ref{f12} is to divide our points in two
shaded areas, one that follows the trend shown by metal poor stars (A1), and the
other that follows the trend shown by metal rich stars (A2). Although the
A2 group is less populated than A1, its presence is definitively proven by our
results, which is one of the most surprising results of this paper.
In both cases the progenitors of each relation appear to be a quite old group
of stars, that for A1 have [Fe/H]$\sim$-2.0 and age$\sim$11.0 Gyrs, while that
for A2 have [Fe/H]$\sim$-1.2 and an age of almost 14 Gyrs. 
Because of their ages and metallicity differences, it is very difficult
that the second derives from the first by some kind of chemical evolution, so they should
be assumed as two independent primordial objects.
Then in each area stars appear to evolve toward higher metallicities following the arrow labeled
with {\it age-metallicity evolution}. We underline the fact that the
age-metallicity evolution we draw is only a first intent of
interpretation. Future works could change the scenario we are 
proposing significantly.

The age-metallicity relation within A1 was already presented in
\citet{hi04} as shown by the blue squares. That within A2 is evidenced here for the
first time due to our much more complete SGB sample. The presence of two
age-metallicity relations can easily explain the age spread of Pop$_4$,
that appears to be very large. It simply comes from the
superposition of the two age-metallicity relations in the range -1.20$<$[Fe/H]$<$-0.90. 

The fact that the oldest population is more metal-rich compared with the bulk
of the metal-poor stars in a globular cluster like object is something that
goes against common sense. One would have expected a monotonically
decreasing age-metallicity relation, or at least that the most metal-poor and
metal-rich stars were coeval, if the initial chemical enrichment would have occured in a
short time-scale ($<$1 Gyr). The only way to rejuvenate A2
stars and place it on the top of A1 is to assume a total CNO $\sim$1.8
dex larger then its actual value, i.e. all star belonging to the A2 group
should have [C+N+O/Fe]$\sim$2.3 dex, that is a much higher value than anyone
observed not only in $\omega$ Cen, but in any Galactic object. Only some very
metal-poor stars \citep{Si06} show extremely enhanced C, N, and O (with
[N/Fe] up to $\sim$+3.0), but in this case this is attributed to contamination by an
AGB or massive fast-rotating companion. 

To complete our interpretation, it is remarkable that each relation 
has an age difference (difference between the youngest and
oldest star, including also the errors) of $\sim$3 Gyrs for a given [Fe/H] value.

At this point it is natural to propose that $\omega$ Centauri is the result
of the merging of two independent objects (dwarf galaxies?) or of two
independent parts of a single larger object, the first having the the oldest stars
more metal poor ([Fe/H]$\sim$-2.0), the second having the oldest stars more
metal rich ([Fe/H]$\sim$-1.2). Each object or part had its own independent evolution in
the age-metallicity plane, at least down to 10 Gyrs. The evolution ended at
$\sim$6 Gyrs for both of them.
If they merged before or after the full evolution of the stars in the
age-metallicity plane, and if and how they interacted, is very hard to say.

\subsection{The impact of the uncertainty of the [C+N+O/Fe] vs. [Fe/H] relation on the
age-metallicity relation}

\begin{figure*}[ht!]
\epsscale{2.0}
\plotone{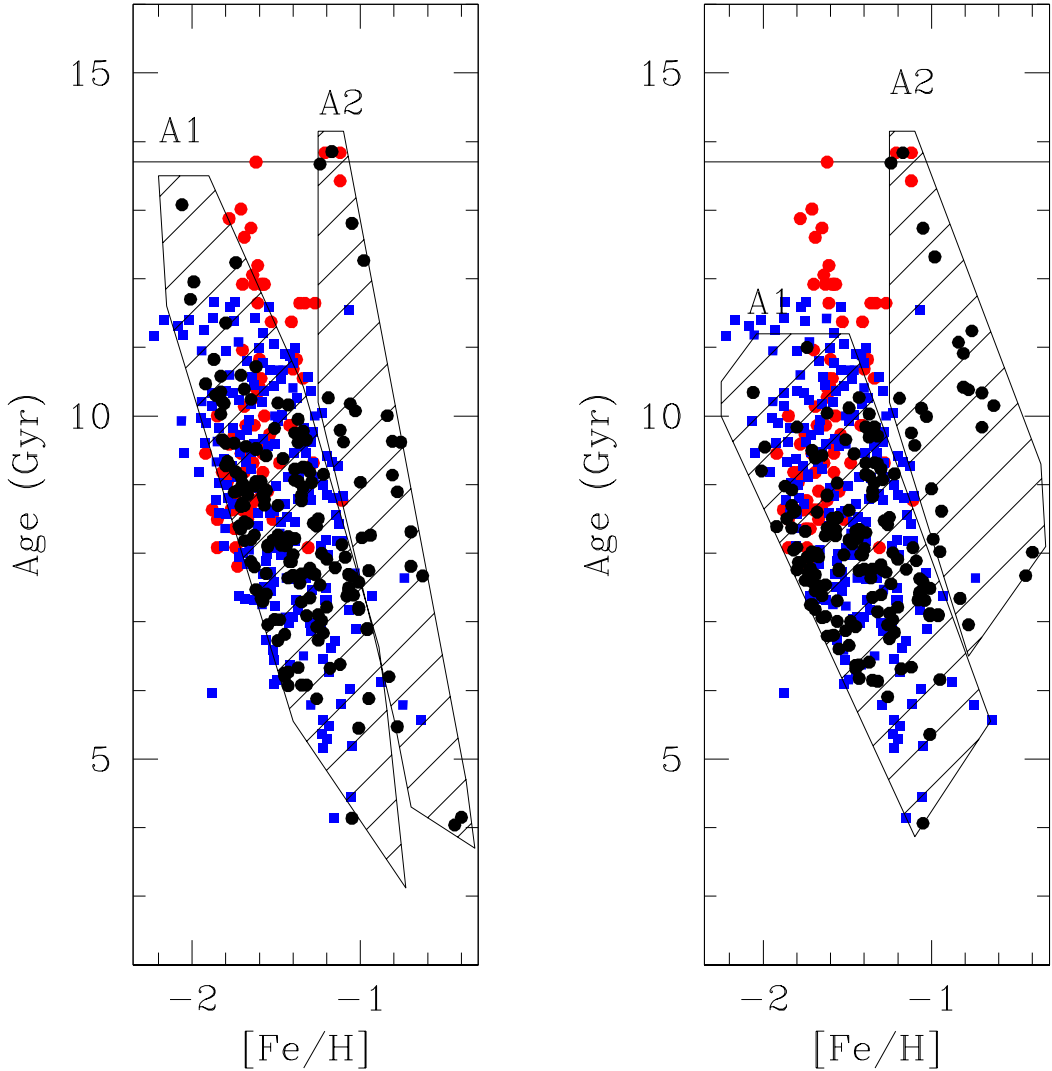}
\caption{The impact of the uncertainty on the [C+N+O/Fe] vs. [Fe/H] relation on the
age-metallicity relation. See text for more details}
\label{f13}
\end{figure*}

As discussed in \citet{ma12}, the total CNO content can heavily influence
the final age of a star and, in our case, the age-metallicity relation.
For this reason we performed the following test in order the check how a incorrect
estimation of the [C+N+O/Fe] trend as a function of [Fe/H] presented in
Fig.~\ref{f11} can alter the final result. We firstly estimated the two most extreme
fits of the data, the first with the highest [C+N+O/Fe] value at
[Fe/H]$\sim$-2.0 and the lowest for [Fe/H]$>$-1.5. This is plotted as a blue
dashed line in Fig.~\ref{f11}. The second with the the lowest [C+N+O/Fe] value at
[Fe/H]$\sim$-2.0 and the highest for [Fe/H]$>$-1.5. This is plotted as a red
dashed line in Fig.~\ref{f11}. We underline that those fits are completely
unreliable, but we want to check if also in the worst case our
conclusions about the age-metallicity relation are supported.
We report the result of this test in Fig.~\ref{f13}. The panel on the left
was obtained with the blue fit, while the one on the right with the red fit.

We see that the presence of the two independent relations A1 and A2 is
confirmed. In the panel on the left the oldest stars in A1 are as old as
the oldest stars in A2, while in the one on the right, oldest
stars in A1 are $\sim$4 Gyrs younger.
This test confirms our hypothesis of the presence of two old, unrelated
population in $\omega$ Centauri with very different iron contents.
On the other hand, each relation shows an age difference of
4 Gyrs (in the first case) or 3 Gyrs (in the second case)
for a given [Fe/H] value. 

To check the impact of the uncertainty in CNO
content on the age-spread for a given [Fe/H], 
we assume that the spread of 0.08 dex we estimated for [C+N+O/Fe] around
the mean trend of Fig.~\ref{f11} is totally intrinsic and not due to measurement
errors. We lack any information about CNO for our stars, so we must proceed in
a statistical way. In the previous section we showed that for a given
metallicity the age difference is $\sim$3 Gyrs. A spread ($\sigma$) of
0.08 dex correspond to a maximum interval in the [C+N+O/Fe] value of $\sim$0.5
dex (i.e. 6$\times$$\sigma$), that translates in an age interval of $\sim$1.5
Gyrs. If we subtract this value to the age difference we are left with
1.5 Gyrs. This means that the age difference is real and larger than
1.5 Gyrs (very likely $\sim$3 Gyrs because the spread of 0.08 dex in the CNO trend is
almost totally due to measurement errors).

\section{Conclusions}

In this paper we analyzed 172 stars belonging to the SGB region of 
$\omega$ Centauri, in order to study the age and metallicity dispersion and the
age-metallicity relation to have further clues of how this object was formed.
For this purpose we obtained medium resolution spectra (R=22500) in the
6120--6405 \AA \ range and measured the iron content of our stars using the
same general approch as \citet{vi07}.

The accuracy of our measurements coupled with the age-sensitivity of the SGB,
allowed us to find that any of the 5 SGBs of the cluster has a distribution in
metallicity with a spread that exceeds the observational errors.
In addition each SGB displays several peaks, that indicate the presence of
several sub-populations. We could identify 6 of them based on their [Fe/H] value.

Taking advantage of the age-sensitivity of the SGB we showed that, first of all,
at least half of the sub-population have an age spread of at least 2 Gyrs. These 
results are indeed very surprising, and we urge confirmation with additional data.

Then, considering all the possible contributors, we tranformed the magnitude
of each star into a relative age, obtaining an age-metallicity relation. 
We underline the fact that we do not use absolute ages but values that consider only the
differential SGB luminosity corrected for the differential He, [Fe/H], and
CNO contents of each star. Because of this our final error on age is
significantly reduced.

Our relation agrees well with those published previously, which however cover
the age-metallicity space only partially.

The interpretation of the age-metallicity relation is not straightforward, but
it is very likely that the cluster (or what we can call its progenitor) was initially composed of two
old populations, but having different
metallicities. The most metal poor had [Fe/H]$\sim$-2.0, while the most metal
rich had [Fe/H]$\sim$-1.2, but the oldest stars in the metal-rich regime
appear to be several Gyrs older than their
oldest metal-poor counterparts. Because of their ages and metallicity, it is very difficult
that the second derives from the first by some kind of chemical evolution, so they should
be assumed as two independent primordial objects. Afterwards, at first
order, each one evolved chemically with iron that linearly increases with
age according to our interpretation. This evolution stopped at $\sim$6
Gyrs. In any case they remain separated in the age-metallicity plane at least down to 10 Gyrs. 
In addition to this, each object shows an age spread of $>$2 Gyrs for a given metallicity.
These conclusions are not altered by any possible uncertainty on the
[C+N+O/Fe] vs. [Fe/H] relation.

These two primordial progenitors could correspond to two dwarf galaxies that at a
given unknown time merged to form what is known today as $\omega$ Centauri.
If they merged before of after the full evolution of the stars in the
age-metallicity plane, and if and how they interacted, it is very hard to say.
Clearly, much further work on this enigmatic object is
required to help solve some of its mysteries.

\acknowledgments

SV and DG gratefully acknowledge support from the Chilean 
BASAL Centro de Excelencia en  Astrofisica y Tecnologias Afines (CATA) grant PFB-06/2007.
S.V. gratefully acknowledges the support provided by FONDECYT N. 1130721.
G.P., R.G.G., and S.C. acknowledge support by INAF 
under contract PRIN INAF 2009 Formation and Early Evolution of Massive Star
Clusters.
SC is grateful for financial support from PRIN-INAF 2011 {\it Multiple Populations in Globular Clusters: their 
role in the Galaxy assembly} (PI: E. Carretta), and from PRIN MIUR 2010-2011, 
project {\it The Chemical and Dynamical Evolution of the Milky Way and Local Group Galaxies}, prot. 2010LY5N2T (PI: F. Matteucci).

\clearpage

\begin{deluxetable}{lcccccccccc}
\tablecolumns{12}
\tablewidth{0pc}
\tablecaption{Parameters of the observed stars. This Table is
published in its entirety in the electronic edition of the Astrophysical Journal.
A portion is shown here for guidance regarding its form and content.}
\tablehead{
\colhead{ID} & \colhead{RA} & \colhead{Dec} & \colhead{$m_{\rm F435W}$} &
\colhead{$m_{\rm F625W}$} & \colhead{T$_{eff}$} & \colhead{log(g)} & \colhead{v$_t$} & 
\colhead{[Fe/H]} & \colhead{RV} & \colhead{Age}
}
\startdata
\hline
            &     Degrees  &     Degrees  &   Mag. &  Mag.  &  K   & dex  & km/s & dex   & km/s &  Gyr \\    
\hline
SGB$_A$.1   & 201.60166667 & -47.55916667 & 18.072 & 16.976 & 5592 & 3.69 & 0.99 & -1.67 & 225.5 &  8.5\\
SGB$_A$.106 & 201.58141667 & -47.44286111 & 18.107 & 17.104 & 5821 & 3.78 & 0.97 & -1.62 & 221.9 &  7.2\\
SGB$_A$.107 & 201.60875000 & -47.44119444 & 18.069 & 16.985 & 5611 & 3.69 & 0.99 & -1.71 & 232.9 &  8.6\\
SGB$_A$.112 & 201.56883333 & -47.43730556 & 18.112 & 17.118 & 5829 & 3.78 & 0.97 & -1.71 & 232.8 &  7.9\\
SGB$_A$.12  & 201.59787500 & -47.54488889 & 18.213 & 17.235 & 5909 & 3.85 & 0.95 & -1.49 & 229.6 &  6.8\\
SGB$_A$.127 & 201.76162500 & -47.42775000 & 18.102 & 17.081 & 5813 & 3.78 & 0.97 & -1.43 & 255.1 &  6.2\\
SGB$_A$.130 & 201.74654167 & -47.42536111 & 18.109 & 17.059 & 5767 & 3.77 & 0.97 & -1.32 & 225.4 &  6.2\\
SGB$_A$.135 & 201.68729167 & -47.42155556 & 18.059 & 16.994 & 5652 & 3.70 & 0.99 & -1.72 & 238.3 &  8.4\\
SGB$_A$.136 & 201.56879167 & -47.42036111 & 18.050 & 16.955 & 5603 & 3.68 & 0.99 & -1.63 & 225.5 &  8.0\\
SGB$_A$.138 & 201.70537500 & -47.41897222 & 18.070 & 16.973 & 5588 & 3.68 & 0.99 & -1.68 & 241.2 &  8.6\\
SGB$_A$.14  & 201.70050000 & -47.54366667 & 18.073 & 17.028 & 5701 & 3.72 & 0.98 & -1.71 & 229.4 &  8.2\\
SGB$_A$.141 & 201.61179167 & -47.41819444 & 18.059 & 17.006 & 5708 & 3.72 & 0.98 & -1.58 & 206.8 &  7.1\\
SGB$_A$.145 & 201.68179167 & -47.41636111 & 18.148 & 17.145 & 5853 & 3.81 & 0.96 & -1.45 & 244.4 &  6.4\\
SGB$_A$.146 & 201.63141667 & -47.41597222 & 18.068 & 17.021 & 5748 & 3.74 & 0.98 & -1.45 & 234.2 &  6.4\\
\hline	   										      
\enddata										      
\label{t2}
\end{deluxetable}

\end{document}